\documentclass[fleqn,10pt]{wlscirep}

\usepackage{graphicx}
\usepackage{hyperref}
\usepackage{color}

\newcommand{\rev}[1]{\textcolor{black}{ #1}}

\title{Impact of origin-destination information in epidemic spreading}

\author[1,*]{Sergio G\'omez}
\author[2]{Alberto Fern\'andez}
\author[3,4]{Sandro Meloni}
\author[1]{Alex Arenas}

\affil[1]{Dept.\ Enginyeria Inform\`atica i Matem\`atiques, Universitat Rovira i Virgili, 43007 Tarragona, Spain}
\affil[2]{Dept.\ Enginyeria Qu\'{\i}\-mica, Universitat Rovira i Virgili, 43007 Tarragona, Spain}
\affil[3]{\rev{IFISC, Institute for Cross-Disciplinary Physics and Complex Systems (CSIC-UIB), 07122 Palma de Mallorca, Spain}}
\affil[4]{BIFI, Instituto de Biocomputaci\'on y F\'{\i}sica de Sistemas Complejos, Universidad de Zaragoza, 50018 Zaragoza, Spain}

\affil[*]{Correspondence and requests for materials should be addressed to S.G.\ (\href{mailto:sergio.gomez@urv.cat}{sergio.gomez@urv.cat})}

\begin{abstract}
The networked structure of contacts shapes the spreading of epidemic processes. Recent advances on network theory have improved our understanding of the epidemic processes at large scale. \rev{The relevance of several considerations still needs to be evaluated in the study of epidemic spreading}. One of \rev{them} is that of accounting for the influence of origin and destination patterns in the flow of the carriers of an epidemic. Here we compute origin-destination patterns compatible with empirical data of coarse grained flows in the air transportation network. We study the incidence of epidemic processes in a metapopulation approach considering different alternatives to the flows prior knowledge. The data-driven scenario where the estimation of origin and destination flows is considered \rev{turns out to be} relevant to assess the impact of the epidemics at a microscopic level (in our scenario, which populations are infected). \rev{However, this information is} irrelevant \rev{to assess its} macroscopic incidence (fraction of infected populations). These results are of interest to implement even better computational platforms to forecast epidemic incidence.
\end{abstract}

\begin{document}

\flushbottom
\maketitle

\thispagestyle{empty}

\section*{Introduction}

The worldwide simulation of diseases spreading is a challenge involving many scales of complexity \cite{anderson1992infectious,hethcote2000mathematics,daley2001epidemic,pastor2015epidemic}. These scales are progressively taken into account by introducing, every time, more realistic elements in the description of the whole scenario. Relatively recent advances on this specific goal came from the consideration of the specific topology of mobility networks that represent main carriers of epidemics at the global scale \cite{pastor2001epidemic,newman2002spread,pastor2015epidemic}. The worldwide airports network has been identified as the fastest interaction mechanism between humans living very far apart. In this context, metapopulation models constitute a natural approach for the analysis of epidemic spreading processes, since they combine in a single framework the local contagions in the so-called reaction phase, and the mobility of the individuals in the diffusion stage \cite{sattenspiel1995structured,hanski1998metapopulation,hufnagel2004forecast,watts2005multiscale,colizza2007reaction,colizza2008epidemic}.

Although metapopulation models have shown its \rev{predictive} power in recent outbreaks, it has been recognized the crucial role played by the mobility patterns \cite{gonzalez2008understanding,balcan2009multiscale,balcan2011phase}. A common hypothesis in metapopulation simulations is that the diffusion through the links of the network is markovian, i.e., each individual willing to move to a neighboring node will choose the destination according to certain fixed probabilities assigned to each of the links \cite{hufnagel2004forecast,colizza2007reaction,colizza2007invasion,colizza2008epidemic}. Usually this method is complemented with memory of the origin node, thus making the individuals come back to their respective residences in a home-to-work travel pattern \cite{balcan2011phase,gomez2017critical}. Another possible extension is the selection of a random destination instead of a neighbor, e.g.\ with probability proportional to the strength of the node, thus allowing for longer trips \cite{meloni2009traffic}. This hypothesis is a shortcut on the difficult problem of assessing how many people travel from a certain origin to a certain destination. Although this information is probably registered by every air carrier, its recollection at world wide scale is not easy. Data scientists usually have at disposal the amounts of (usually annual) accumulated number of passengers between two connected places, that is the weight of the link, but no information about initial origin or final destination is provided.

In this work we present a methodology to assess origin-destination flows in the air transportation networks, and after we analyze the results obtained with these more realistic flows of passengers on the incidence of epidemic spreading considering metapopulation models \cite{sattenspiel1995structured,watts2005multiscale}. Our results reveal that the consideration of real flow patterns affect the microscopic description of the incidence at the level of cities, while the statistical aggregated incidence is similar to those predicted without considering origin-destination flows.

Essentially, we could summarize that the hard problem of estimating origin-destination patterns can be avoided if the interest is focused on macroscopic details of the epidemic spreading, however it should be considered when microscopic details about the epidemic spreading are needed. In the following we present the method used to estimate the data-driven origin-destination patterns, the metapopulation model, and the comparison of the different routing flow strategies compared to the data-driven scenario. To conclude we present a discussion about the convenience of using origin-destination data-driven flows in the assessment of epidemic spreading.

\section*{Results}

\subsection*{Estimation of an Origin-Destination matrix}

For most kind of analyses in transportation networks, there is a need for origin-destination (O-D) matrices, which specify the travel demands between the origin and destination nodes in the network. Here we wonder up to which point the estimation of O-D matrices are essential factors to determine the outcome of a certain epidemic process that uses the transportation network as the substrate for the carriers. Let us first determine what is the mathematical problem we are facing.

Given a transport network with $n$ nodes and $m$ links, an Origin-Destination (O-D) matrix $T$ is a two-dimensional trip table whose entries $t_{ij}$ represent the number of trips going from origin node $i$ to destination node $j$. Let $v^{a}$ be the observed trip volume on link $a$ of the network, and let $p_{ij}^{a}$ be the proportion of trips going from origin node $i$ to destination node $j$ that use link $a$. For the latter proportions we assume that travelers follow shortest-path routes and, in case of several alternatives, any of them is selected at random with equal probability.

The base constraints to be satisfied for an O-D matrix estimated from link counts state that the sum of all the trips crossing a given link must be equal to the link counts observed on that link,
\begin{equation}
  \sum_{i=1}^{n} \sum_{j=1}^{n} p_{ij}^{a} t_{ij} = v^{a}, \qquad a = 1, \ldots, m,
\label{constraints}
\end{equation}
and we have to add the lower bounds for the number of trips,
\begin{equation}
  t_{ij} \geq 0, \qquad i,j = 1, \ldots, n.
\label{bounds}
\end{equation}

The system of equations formed by Eqs.~(\ref{constraints}) and~(\ref{bounds}) has always more unknowns to estimate ($n \times n$ O-D cells) than base constraints ($m$ links), the only exception being the complete networks (cliques); in fact, most real networks are sparse, thus the number of unknows is much larger than the number of equations, $n^2\gg m$. Given the possibility of existence of multiple solutions, additional considerations to select a preferred O-D matrix are needed. Several approaches have been traditionally used to solve this problem, from which we have selected a maximum entropy model that estimates the most likely trip matrix consistent with observed link counts. The application of the entropy maximization principles to the O-D estimation problem was initially proposed by Willumsen \cite{willumsen1978estimation} and Van Zuylen \cite{van1980most}. In the maximum entropy approach, the most likely trip matrix is the one having the greatest number of microstates associated with it, what is equivalent to estimate an O-D matrix that adds as little information as possible to the knowledge contained in the link counts.

Let $t$ be the total number of O-D trips traversing a network. Then the number of ways, defined as entropy, in which $t$ trips can be divided into groups of $t_{ij}$ trips can be computed as
\begin{equation}
  S = \frac{t!}{\displaystyle \prod_{i=1}^{n} \prod_{j=1}^{n} t_{ij}!}.
\label{objective_1}
\end{equation}
To be able to compute the derivatives needed to optimize the objective function (\ref{objective_1}), we take first the natural logarithm of the function and make use of Stirling's approximation of the factorials, $ \ln(t!) \approx t \ln(t) - t $, thus obtaining a more computationally convenient objective function:
\begin{equation}
  \ln(S) \approx t \ln(t) - t - \sum_{i=1}^{n} \sum_{j=1}^{n} \left( t_{ij} \ln(t_{ij}) - t_{ij} \right).
\label{objective_2}
\end{equation}
Assuming that $t$ is constant and changing the sign of the function, the first two terms can be dropped and the goal becomes to minimize the function
\begin{equation}
  E = \sum_{i=1}^{n} \sum_{j=1}^{n} \left( t_{ij} \ln(t_{ij}) - t_{ij} \right)
\label{objective_3}
\end{equation}
subject to the constraints in Eqs.~(\ref{constraints}) and~(\ref{bounds}).

\subsection*{Epidemic spreading model}

The epidemic spreading model we use is based on the well-known reaction-diffusion \cite{colizza2007reaction} and metapopulation \cite{colizza2008epidemic} models, but incorporating a routing scheme for the moving individuals similar to those in \cite{meloni2009traffic,meloni2011modeling}. The main ingredients are the sites or nodes (e.g.\ cities or airports), where the individuals of the global population are located, and the links between these nodes, which represent the communication channels (e.g.\ roads or airways). Each link has an associated weight which accounts for the traffic, the number of individuals moving between nodes in a certain amount of time. The contagion model we have chosen is the standard susceptible-infected-removed (SIR) one, a good proxy for diseases in which the individuals acquire immunization after being infected. In this model individuals may be in three different states: infected (I), those who have got the disease; susceptible (S), healthy but which may become ill by contacts with infected individuals; and removed (R), once they have recovered from the disease and become immune to it. The dynamics of the epidemic spreading consists of two alternating phases, a reaction in which the individuals in the nodes' subpopulations merge, and a diffusion of some individuals through the communication channels. In the reaction step the subpopulation is considered as well-mixed, i.e., every individual is in contact with the rest in the same node, and here is where the contagions and recoveries take place. There are no contagions between individuals at different nodes, but the disease spreads due to the mobility in the diffusion step.

In the standard metapopulation model \cite{colizza2008epidemic} the diffusion is just a Markovian process, where individuals move to neighboring nodes with a probability proportional to the weight of the link; we call this a {\em Random Diffusion} (RD) process. A more realistic mobility is obtained if every individual has a ``home'' node, it travels to a selected destination, and finally comes back home after spending some time in the destination. It would be possible to choose uniformly {\em Random Destinations} (RU), however it is more plausible a selection proportional to the strength of each possible target subpopulation as in \cite{meloni2011modeling}, what we call here the {\em Strength Proportional} (SP) scheme. Finally, the information in the {\em O-D matrix} defines a new diffusion model (OD), since it directly tells the number of trips from any origin to any destination.

In all the diffusion models where the destinations are not bound to be neighbors of the origin, it is necessary to establish the paths followed by the travelers, i.e., a routing algorithm. We have decided just to select the shortest path (in number of hops) and, if it is not unique, one of them is chosen at random with equal probability. It is possible to improve this routing algorithm taking into account geographic and economic constraints that restrict the feasible available paths, e.g.\ travel times and costs; however, this would only affect the availability and priority of paths, not the results and conclusions from this work.

\subsection*{Microscopic analysis}

First, we analyze the results obtained from our metapopulation spreading models at the microscopic level of nodes and links. We make use of the World Air Transportation Network (ATN) \cite{guimera2005worldwide}, considered as a directed and weighted network. For the calculation of its O-D~matrix, we have only considered trips of length lower or equal to three, both for computational reasons and also because longer trips are rarely observed in real life (see Methods). All the presented results come from averages over $150$ randomly chosen initial conditions in the Monte Carlo simulations. In all the cases, the mobility parameters have been chosen to ensure that at every time step the average number of circulating individuals is the same for all the mobility strategies. See Methods for a full description of the implementation details of our metapopulation epidemic spreading dynamics.

\begin{figure}[tb!]
  \begin{center}
      \includegraphics[width=0.95\columnwidth]{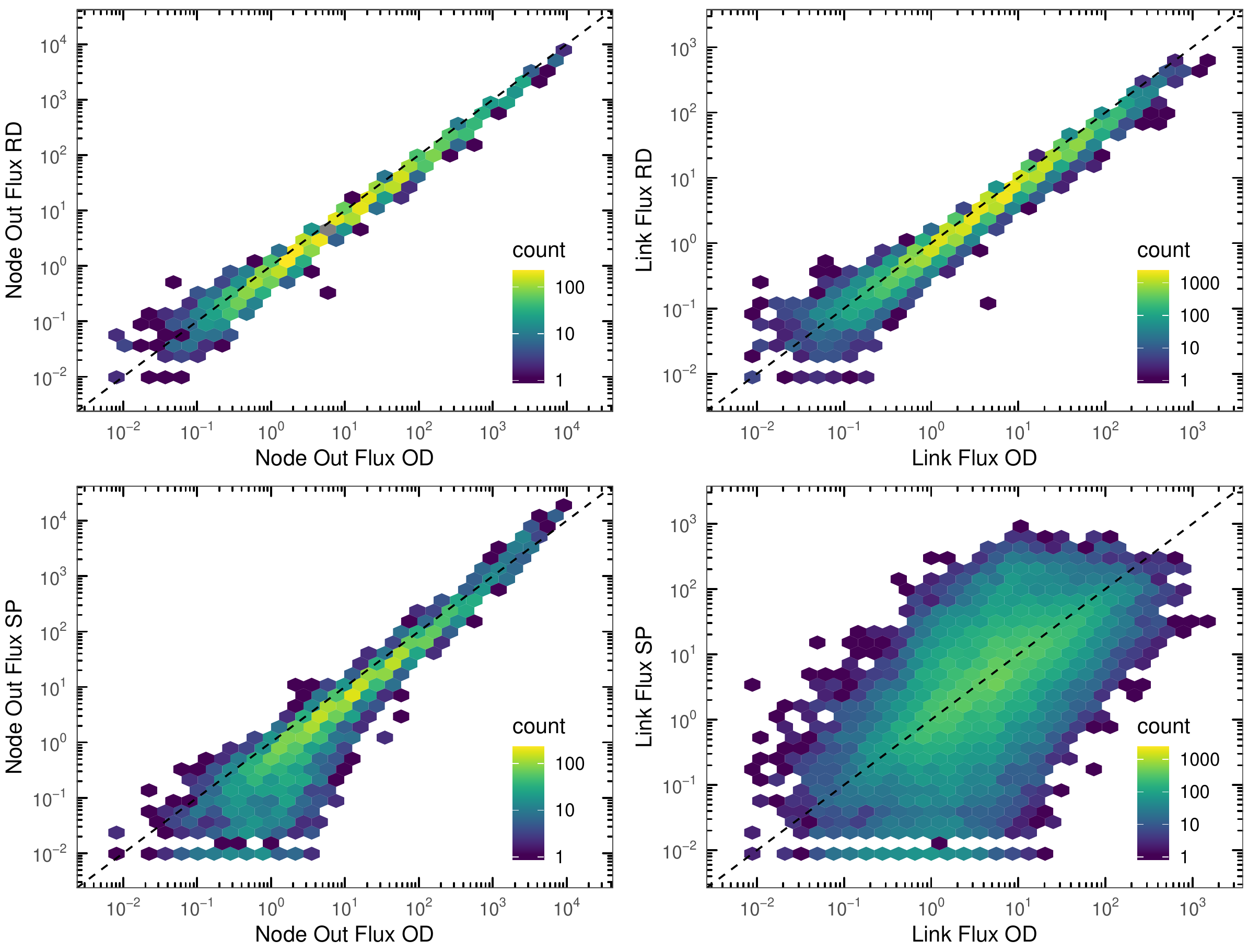}
  \end{center}
  \caption{\textbf{Node and link output fluxes.} \rev{Histogram} of the registered outgoing fluxes at node level (left) and at link level (right), comparing the two strategies Random Diffusion (top) and Strength Proportional (bottom) \rev{with respect to} the O-D strategy. \rev{The color accounts for the number of nodes (left) or links (right) in each bin of the histogram. The values of the Pearson correlation are $0.998$ (RD) and $0.952$ (SP) for node flux, and $0.979$ (RD) and $0.304$ (SP) for link flux.} The network is the World ATN, and the measured normalized fluxes do not depend on the details of the epidemic process. \label{micro_flowout_od}}
\end{figure}

In Fig.~\ref{micro_flowout_od} we plot the outgoing fluxes obtained with Random Diffusion and Strength Proportional schemes, comparing them to the ones obtained with O-D flows. At the level of nodes, both the Random Diffusion and the Strength Proportional strategies show fluxes clearly related to the O-D strategy ones. However, when we take into consideration the link fluxes, we can see that the behavior of the two strategies is different. In the case of the Random Diffusion process, link fluxes also show a linear relation with the O-D fluxes. On the contrary, the link fluxes obtained with the Strength Proportional strategy are not related at all with the O-D ones. The different behavior between the two strategies is explained because both the O-D and the Random Diffusion strategies satisfy constraints~(\ref{constraints}): the O-D matrix satisfies them explicitly as a result of a constrained optimization problem, and the Random Diffusion strategy satisfies them implicitly by the Markovian diffusion process that takes into account the weights of the links. On the contrary, the routing strategy used in the Strength Proportional approach sends travelers through shortest paths without taking into account the weights of the links. Therefore, according to the observed fluxes, we can conclude that travelers pass through the same airports independently of the diffusion strategy used, albeit the routes followed are completely different.

\begin{figure}[tb!]
  \begin{center}
      \includegraphics[width=0.85\columnwidth]{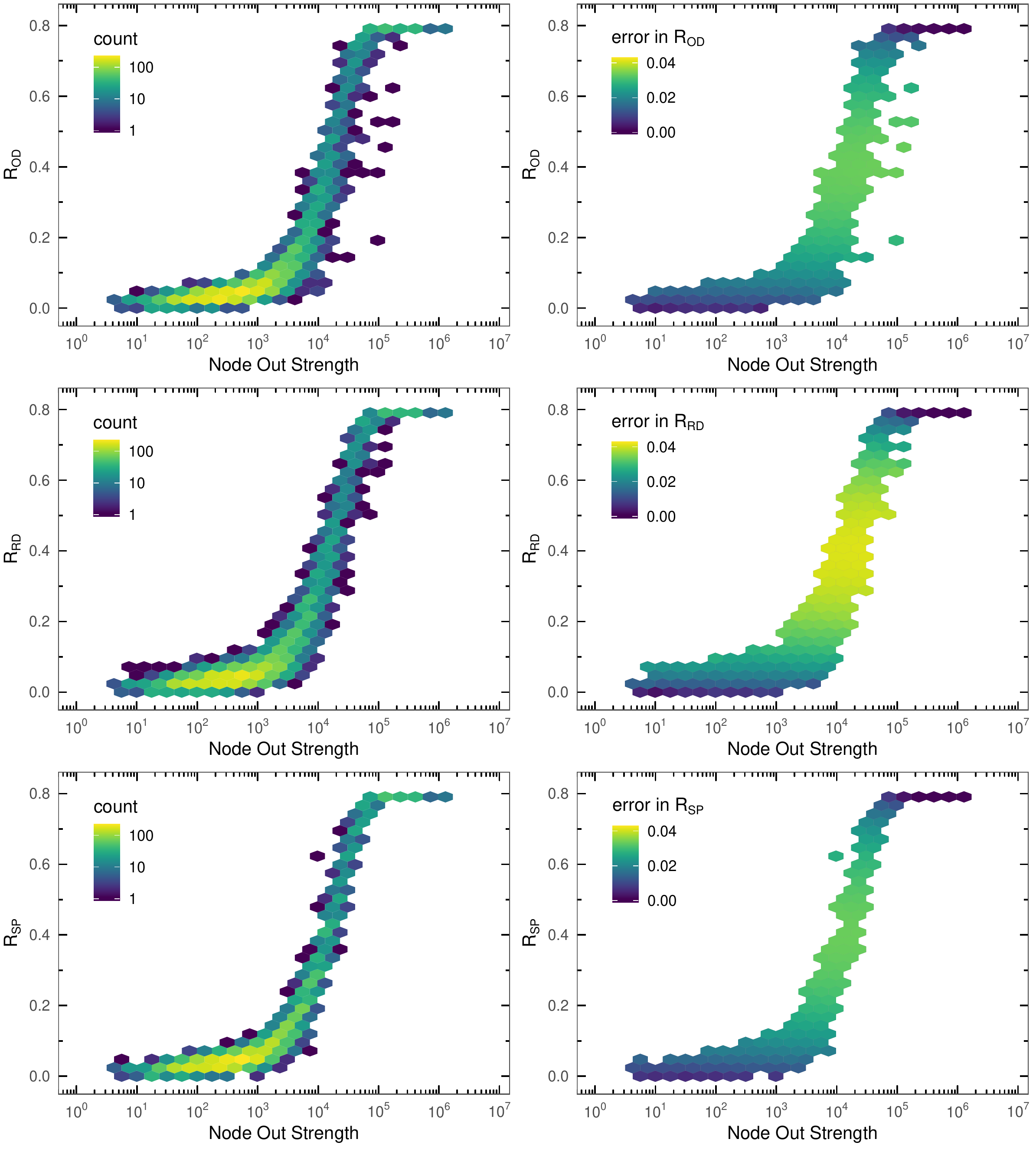}
  \end{center}
  \caption{\textbf{Incidence of the epidemics on the nodes.} \rev{Left:} distribution of the incidence of the epidemics $R$ across the nodes, as a function of the out strength of the node, for the different routing schemes, O-D strategy (top), Random Diffusion (middle), and Strength Proportional (bottom). \rev{The value of the incidence for each node has been} obtained by averaging over $150$~Monte Carlo simulations. \rev{Right:} the corresponding estimated errors on the previous mean values. \rev{The color encodes the number of nodes in each bin (left) and the average error in the epidemic incidence for the nodes in each bin (right).} The network is the World ATN, and the parameters of epidemic spreading process are $\mu=0.04$, $R_0=2.0$, and $\lambda=5 \cdot 10^{-6}$. \label{epidemic_incidence}}
\end{figure}

\begin{figure}[tb!]
  \begin{center}
      \includegraphics[width=0.95\columnwidth]{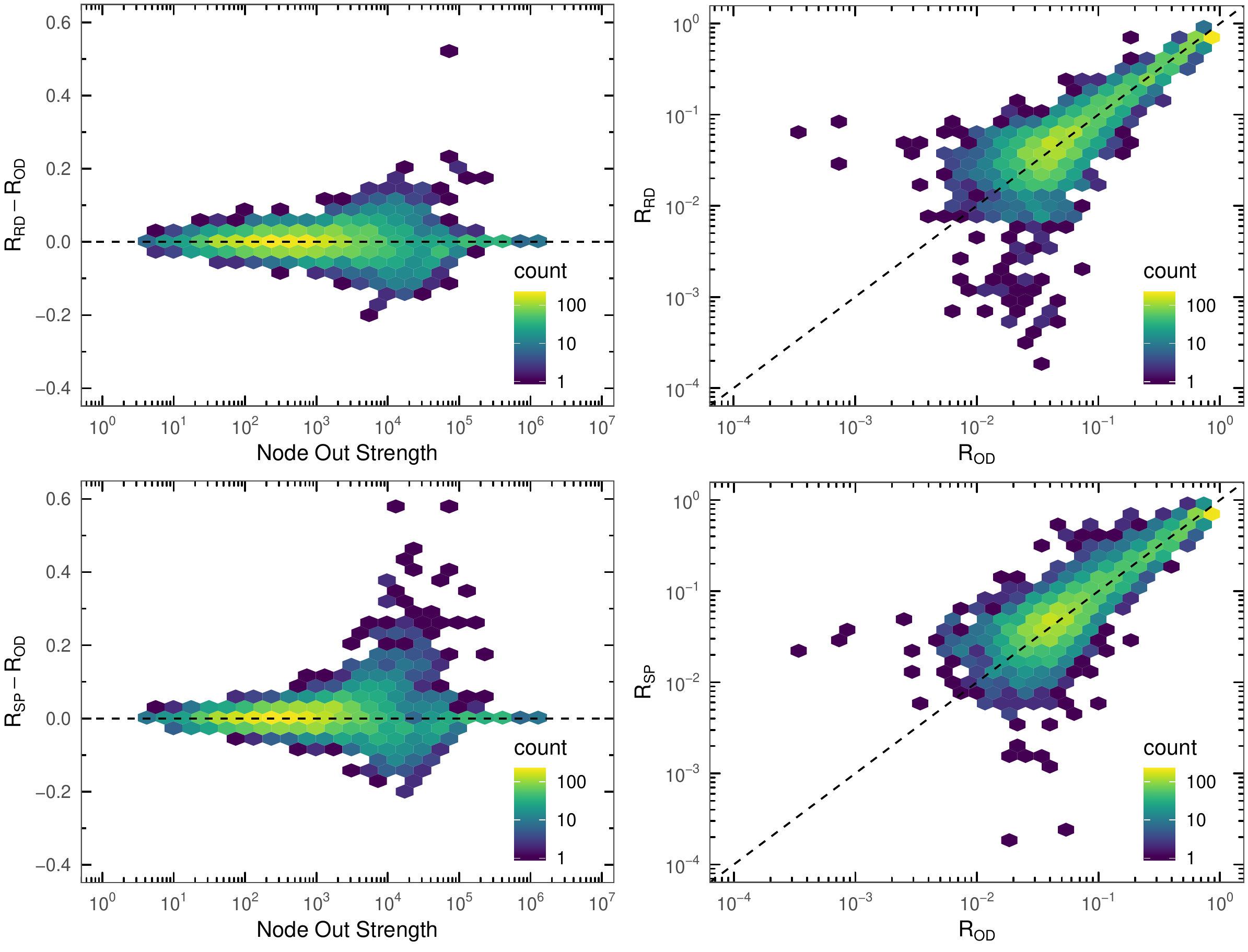}
  \end{center}
  \caption{\textbf{Comparison of the incidence of the epidemics on the nodes.} Left: Differences in the fraction of recovered individuals per node, as a function of the out strength of the node, comparing the two strategies, Random Diffusion (top) and Strength Proportional (bottom), \rev{with respect to} the O-D strategy. Right: \rev{Histogram} of the fraction of recovered individuals $R$ at $t \rightarrow \infty$ comparing the two strategies, Random Diffusion (top) and Strength Proportional (bottom), \rev{with respect to} the O-D strategy. The network is the World ATN, and the parameters of epidemic spreading process are $\mu=0.04$, $R_0=2.0$, and $\lambda=5 \cdot 10^{-6}$. \label{micro_nodes}}
\end{figure}

Although the Random Diffusion and the Strength Proportional strategies show different routing dynamics, it should be analyzed whether these differences are relevant for the epidemic spreading process. With this purpose, we compute the fraction of recovered individuals $R$ at each node (remember that a node is a population) for both diffusion strategies, and we compare them to the ones obtained with the O-D flows. We can see in Fig.~\ref{epidemic_incidence} the distribution of $R$ for the three mobility schemes. \rev{We observe that,} the larger the population size in the node (which is proportional to its strength, see Methods), the larger the incidence of the epidemics, with a non-linear dependency and an important dispersion between nodes of similar strength. The three schemes show a similar distribution profile, thus we need a more direct comparison of incidences node to node. Since the estimated errors on the mean incidence values at each node are \rev{mostly below~$0.03$ (see Fig.~\ref{epidemic_incidence})}, differences in the incidence about or larger than~$0.1$ (in absolute value) can be considered as statistically significant. In the left plots of Fig.~\ref{micro_nodes} we can observe an increasing difference as a function of the out strength of the node up to values about $10^4$, and then the differences start to decrease. The right plots of Fig.~\ref{micro_nodes} show the fraction of recovered individuals $R$ from the Random Diffusion and the Strength Proportional strategies comparing them \rev{with respect to} the ones obtained with the O-D flows. Again, we do not observe any difference between the patterns followed by the Random Diffusion and the Strength Proportional approaches, but it has become evident that the routing strategies greatly affect the epidemic spreading at the level of nodes. Therefore, the availability of O-D information is completely necessary to be able to have good predictions of the microscopic incidence of the epidemics.

\subsection*{Macroscopic analysis}

According to the results presented in Figs.~\ref{micro_flowout_od} and~\ref{micro_nodes}, the different routes that travelers follow in the Random Diffusion and the Strength Proportional strategies show the same patterns of discrepancies when we compare them to the O-D flows in epidemic spreading terms at the microscopic level. Next, we analyze whether these microscopic differences have any significant effect at the macroscopic level.

\begin{figure}[tbp!]
  \begin{center}
      \includegraphics[width=0.7\columnwidth]{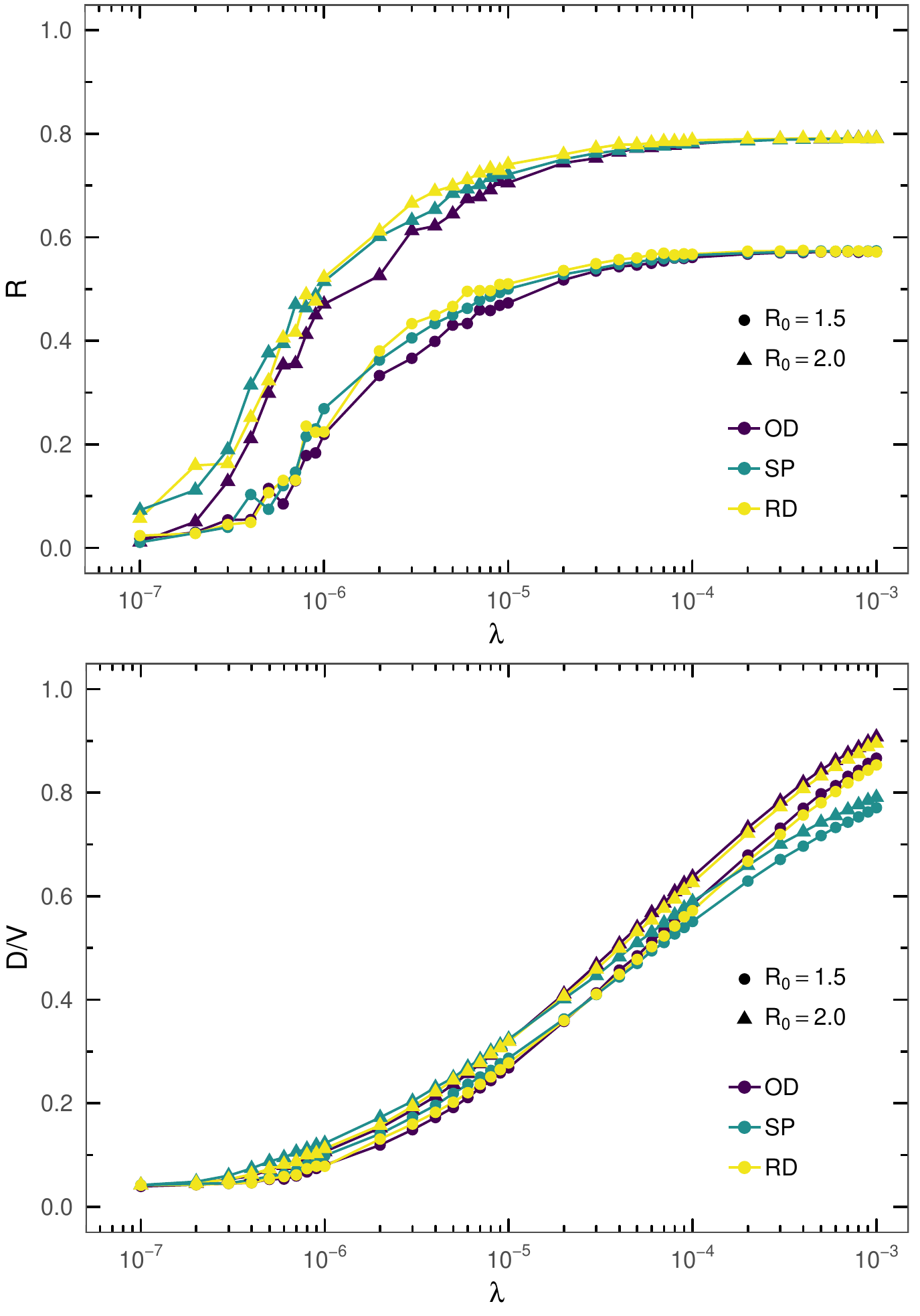}
  \end{center}
  \caption{\textbf{Global incidence of the epidemics for varying mobility.} Fraction of recovered individuals $R$ (top) and fraction of subpopulations that experienced an outbreak $D/V$ (bottom) as a function of the traffic generation rate $\lambda$. The network is the World ATN subnetwork, and the epidemic spreading process has a recovery rate $\mu=0.04$. \label{atn_lambda}}
\end{figure}

In Fig.~\ref{atn_lambda} we show both the fraction of recovered individuals $R$ and the fraction of subpopulations that experienced and outbreak $D/V$ as a function of the traffic generation rate $\lambda$. The results show that, especially for the fraction of recovered individuals, the deviations observed between the O-D matrices and the Random Diffusion and Strength Proportional strategies are minimal at the macroscopic level. However, the Strength Proportional strategy seems to predict a slightly lower value of $D/V$ than the other two strategies, with a deviation which increases for high values of the mobility.

\begin{figure}[tbp!]
  \begin{center}
    \includegraphics[width=0.7\columnwidth]{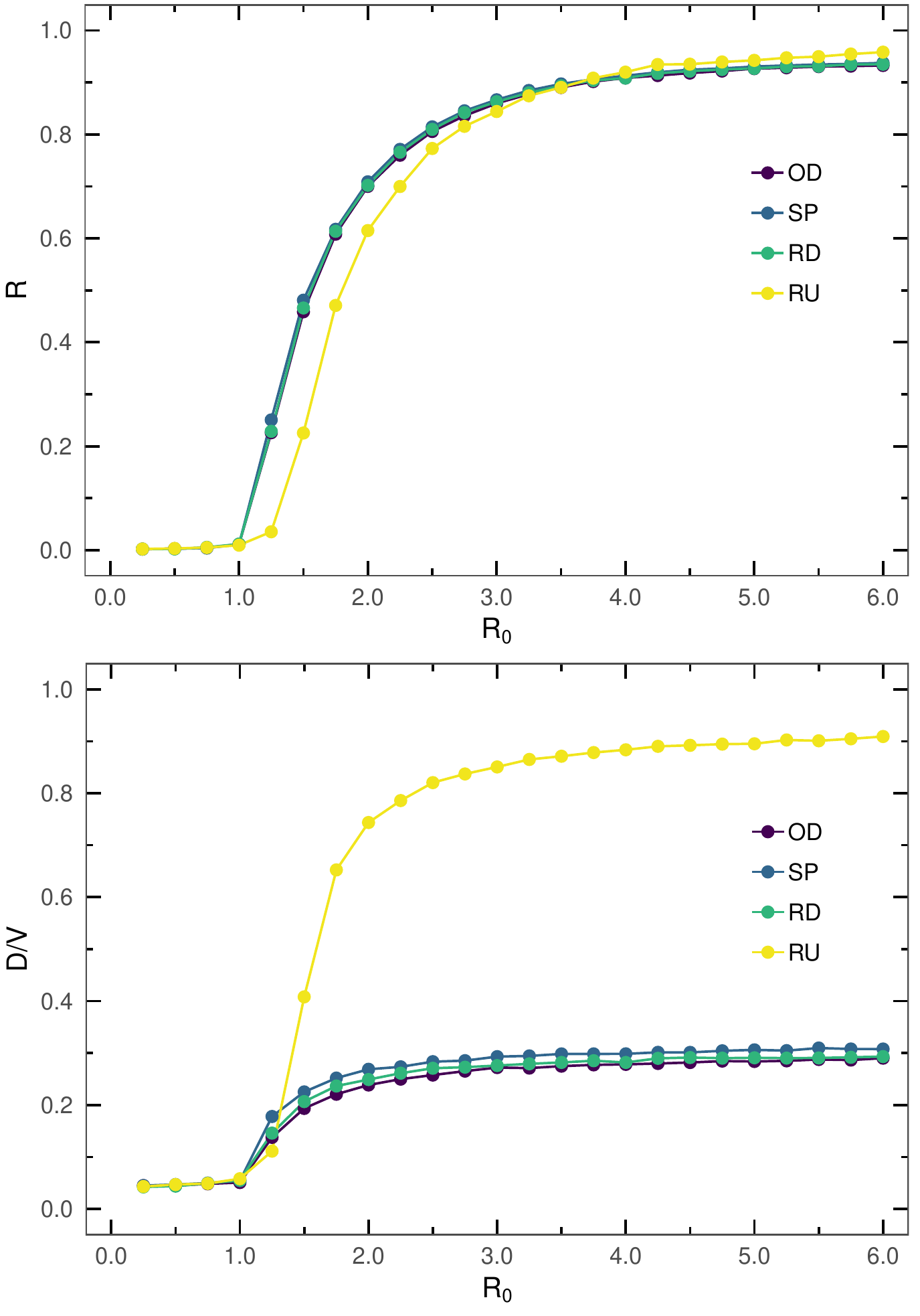}
  \end{center}
  \caption{\textbf{Global incidence of the epidemics for varying reproductive number.} Fraction of recovered individuals $R$ (top) and fraction of subpopulations that experienced an outbreak $D/V$ (bottom) as a function of the reproductive number $R_0$ for different routing strategies. The network is the World ATN, and the parameters of epidemic spreading process are $\mu=0.04$ and $\lambda=5 \cdot 10^{-6}$. \label{world_beta}}
\end{figure}

Analyzing the effect of the reproductive number at the macroscopic level gives similar results. In Fig.~\ref{world_beta} we show the fraction of recovered individuals $R$ and the fraction of subpopulations that experienced and outbreak $D/V$ as a function of the reproductive number $R_0$. Taken together the results of Figs.~\ref{atn_lambda} and~\ref{world_beta}, they confirm that only minimal differences are present between the three diffusion strategies in terms of epidemic spreading at the macroscopic level. In the same figure we also show that this behavior observed for the Random Diffusion and the Strength Proportional strategies cannot be generalized to the Random Destination strategy, which does not take into account the weights of the links in the network. The Random Destination strategy shows a slight reduction in the fraction of recovered individuals, but an enormous overestimation of the fraction of infected subpopulations, thus discarding it as a feasible mobility method.

\section*{Discussion}

We have analyzed the influence of the three different mobility models in a metapopulation model of epidemic spreading. Two of them, the Random Diffusion and Strength Proportional routing strategies, have been traditionally used in the simulation of epidemic spreading processes. Here, we have compared them against the more realistic O-D matrix scheme, which is able to capture not only the observed fluxes through the links of the network, but also the fraction of individuals moving between each origin-destination pair. We have shown how the origin-destination matrix can be calculated, and the differences between using these three different models.

Although the three methods show equivalent fluxes per node, the Strength Proportional clearly deviates at the level of fluxes per link, thus indicating the routes that travelers follow are completely different. These differences have an important impact on the epidemic spreading at the microscopic level of nodes, with significant departures in the incidence of the epidemics in the subpopulations at each node, and being especially important for nodes with intermediate values of the output strength. The observed deviations affect both the Strength Proportional and Random Diffusion models when compared with the more reasonable O-D matrix scheme, and with a very similar deviation pattern.

When we consider the epidemic spreading at the macroscopic level of the whole World ATN network, we realize the differences almost vanish, thus making the Random Diffusion, the Strength Proportional and the O-D matrix approaches basically equivalent. This means it is safe to use any of the three mobility models if the interest just lays on the global incidence of the epidemics, but care must be taken when the details are needed at the level of nodes, where only origin-destination information is able to provide the desired quality of the predictions.

\section*{Methods}

\subsection*{O-D matrix of air transportation network}

The World Air Transportation Network (ATN) data set used in this work is composed of passenger flights operating between November 1, 2000 and October 31, 2001 as compiled by the OAG Worldwide (Downers Grove, IL) and analyzed previously in \cite{guimera2005worldwide}. The network is formed by 3618~different airports, that means the existence of a total of 13086306 possible O-D trips. The analysis of all the shortest paths using the Floyd-Warshall algorithm \cite{floyd1962algorithm} shows that there are 27028 (0.21\%) direct connections (shortest paths of length one), 440038 (3.36\%) connections with one transfer (length two), and 2463230 (18.82\%) with two transfers (length three). These are the only trips considered here, since longer air trips are rarely observed in real life. Additional cost factors $c_{ij}$ (equal to 1, 2, or 3), accounting for the length of the route from origin node $i$ to destination node $j$, have been added to the objective function $E$ in Eq.~(\ref{objective_3}) in order to avoid the dominance of longer trips:
\begin{equation}
  E_c = \sum_{i=1}^{n} \sum_{j=1}^{n} c_{ij} \left( t_{ij} \ln(t_{ij}) - t_{ij} \right).
  \label{objective_4}
\end{equation}
With this final formulation of the objective function $E_c$, the O-D matrix estimation problem becomes a separable convex optimization problem that we have solved using MOSEK \cite{mosek2018mosek}, a software especially designed for large-scale mathematical optimization problems.

\subsection*{Epidemic spreading model implementation}

The epidemic model is implemented as follows. Each simulation starts with a small fraction of infected individuals, $I_0$. Namely, we randomly choose a small fraction of sites (less than 1\%) and within these subpopulations only the 1\% of the individuals is infected, ensuring that the condition $I_0 > \frac{\mu}{\beta}$ is fulfilled, where $\beta$ and $\mu$ are the infection and recovery rates, respectively, and $R_0 = \frac{\beta}{\mu}$ is the so-called reproductive number. In the simulations the diffusion and reaction dynamics have the same time scale so, at each time step, first a movement step is performed and then the SIR dynamics takes place.

In the diffusive phase, for each possible pair of origin-destination nodes $(i,j)$, the number of individuals starting a trip from node $i$ towards node $j$ is given by a binomial distribution with parameters the  subpopulation size $N_i(t)$ and the probability $p_{ij}=\lambda\,\frac{t_{ij}}{\sum_k t_{ik}}$, where $0 \leq \lambda \leq 1$ is a mobility rate to distinguish between different traffic regimes, and $t_{ij}$ is the $(i,j)$-th entry of the O-D matrix. To simulate the Strength Proportional and Random Destinations schemes, the O-D matrix is substituted respectively by a Strength-Destination matrix $S$ whose elements are equal to the input strength of the destination node, $s_{ij} = w_{j}^{\mbox{\scriptsize(in)}} = \sum_k w_{kj}$, and by an uniform Random-Destination matrix $U$ in which all the elements equal the unity, $u_{ij} = 1$. Once the number of traveling individuals and their respective destinations have been chosen, they start a shortest path trip to their destination, making one hop to a neighboring node at each time step. When the individuals arrive to their destinations, they come back to their home nodes, also following a shortest path, with one hop per time step. This travel pattern yields and almost stationary size of the subpopulations for all the mobility schemes, provided the mobility rate is not too large.

After the diffusive step, the reaction dynamics is evaluated. As a proxy of the initial population of each node $N_i(0)$ we use node's strength $w_i$ calculated in the original ATN data \cite{guimera2005worldwide}, which amounts a total of $5.82\cdot 10^7$ individuals in the network. The distribution of subpopulation sizes (i.e., strengths) approximately follows a power-law distribution, with sizes ranging between~10 and $3.05\cdot 10^6$~individuals. When time goes on, $N_i(t)$ changes according to the chosen destination selection scheme and routing strategy. Within the nodes, one step of a SIR process takes place, supposing a well-mixed population. The state of every individual inside a node $i$ is modified according to the following probabilities: a susceptible individual becomes infected with probability $p^{(S\rightarrow I)} = 1 - (1- \frac{\beta}{N_i})^{I_i}$, and an infected one recovers with probability $p^{(I\rightarrow R)}=\mu$. Specifically, the exact number of individuals that change their state is determined by binomial distributions with probabilities $p^{(S\rightarrow I)}$ and $p^{(I\rightarrow R)}$, and population sizes $S_i(t)$ and $I_i(t)$ of susceptible and infected individuals, respectively. Note that in this scenario, $R_0$ only participates in the internal nodes' dynamics; individuals traveling through node $i$ are involved in the epidemic dynamics and thus they can change their state while at node $i$. Finally, simulations end when the stationary state $I(t)=0$ is reached.


\section*{Acknowledgements}

This work has been partially supported by MINECO through grant FIS2015-71582-C2-1 (A.A.\ and S.G.), Generalitat de Catalunya project 2017-SGR-896 (A.A.\ and S.G.), and Universitat Rovira i Virgili projects 2017PFR-URV-B2-41 (A.A.\ and S.G.) and 2017PFR-URV-B2-29 (A.F.). S.M.\ is supported by MINECO through the Ram\'on y Cajal program. A.A.\ acknowledges financial support from ICREA Academia and the James S.\ McDonnell Foundation.

\section*{Author contributions statement}

All authors designed the research, analyzed the results and wrote the manuscript. A.F.\ and S.M.\ performed the simulations.

\section*{Additional information}

\textbf{Competing interests:} The authors declare no competing financial and/or non-financial interests.

\newpage
\section*{Figure legends}

\noindent
\textbf{Figure 1. Node and link output fluxes.} \rev{Histogram} of the registered outgoing fluxes at node level (left) and at link level (right), comparing the two strategies Random Diffusion (top) and Strength Proportional (bottom) \rev{with respect to} the O-D strategy. \rev{The color accounts for the number of nodes (left) or links (right) in each bin of the histogram. The values of the Pearson correlation are $0.998$ (RD) and $0.952$ (SP) for node flux, and $0.979$ (RD) and $0.304$ (SP) for link flux.} The network is the World ATN, and the measured normalized fluxes do not depend on the details of the epidemic process.
\\

\noindent
\textbf{Figure 2. Incidence of the epidemics on the nodes.} \rev{Left:} distribution of the incidence of the epidemics $R$ across the nodes, as a function of the out strength of the node, for the different routing schemes, O-D strategy (top), Random Diffusion (middle), and Strength Proportional (bottom). \rev{The value of the incidence for each node has been} obtained by averaging over $150$~Monte Carlo simulations. \rev{Right:} the corresponding estimated errors on the previous mean values. \rev{The color encodes the number of nodes in each bin (left) and the average error in the epidemic incidence for the nodes in each bin (right).} The network is the World ATN, and the parameters of epidemic spreading process are $\mu=0.04$, $R_0=2.0$, and $\lambda=5 \cdot 10^{-6}$.
\\

\noindent
\textbf{Figure 3. Comparison of the incidence of the epidemics on the nodes.} Left: Differences in the fraction of recovered individuals per node, as a function of the out strength of the node, comparing the two strategies, Random Diffusion (top) and Strength Proportional (bottom), \rev{with respect to} the O-D strategy. Right: \rev{Histogram} of the fraction of recovered individuals $R$ at $t \rightarrow \infty$ comparing the two strategies, Random Diffusion (top) and Strength Proportional (bottom), \rev{with respect to} the O-D strategy. The network is the World ATN, and the parameters of epidemic spreading process are $\mu=0.04$, $R_0=2.0$, and $\lambda=5 \cdot 10^{-6}$.
\\

\noindent
\textbf{Figure 4. Global incidence of the epidemics for varying mobility.} Fraction of recovered individuals $R$ (top) and fraction of subpopulations that experienced an outbreak $D/V$ (bottom) as a function of the traffic generation rate $\lambda$. The network is the World ATN subnetwork, and the epidemic spreading process has a recovery rate $\mu=0.04$.
\\

\noindent
\textbf{Figure 5. Global incidence of the epidemics for varying reproductive number.} Fraction of recovered individuals $R$ (top) and fraction of subpopulations that experienced an outbreak $D/V$ (bottom) as a function of the reproductive number $R_0$ for different routing strategies. The network is the World ATN, and the parameters of epidemic spreading process are $\mu=0.04$ and $\lambda=5 \cdot 10^{-6}$.

\end{document}